\documentclass[twocolumn,prx,superscriptaddress,showpacs,nobibnotes]{revtex4-1}

\usepackage{amsmath}

\usepackage{graphicx}
\usepackage{lipsum}
\usepackage{hyperref}
\usepackage{color}

\begin{document}

\title{Dynamical spin susceptibility in La$_{2}$CuO$_4$ studied by resonant inelastic x-ray scattering}

\author{H. C. Robarts} 
\affiliation{Diamond Light Source, Harwell Campus, Didcot OX11 0DE, United Kingdom}
\affiliation{H. H. Wills Physics Laboratory, University of Bristol, Bristol BS8 1TL, United Kingdom}

\author{M. Garc\'ia-Fern\'andez}
\affiliation{Diamond Light Source, Harwell Campus, Didcot OX11 0DE, United Kingdom}

\author{J.\;Li}
\affiliation{Diamond Light Source, Harwell Campus, Didcot OX11 0DE, United Kingdom}
\affiliation{Beijing National Laboratory for Condensed Matter Physics and Institute of Physics, Chinese Academy of Sciences, Beijing 100190, China}

\author{A. Nag}
\affiliation{Diamond Light Source, Harwell Campus, Didcot OX11 0DE, United Kingdom}

\author{A. C. Walters}
\affiliation{Diamond Light Source, Harwell Campus, Didcot OX11 0DE, United Kingdom}

\author{N. E. Headings}
\affiliation{H. H. Wills Physics Laboratory, University of Bristol, Bristol BS8 1TL, United Kingdom}

\author{S. M. Hayden}
\email{s.hayden@bristol.ac.uk}
\affiliation{H. H. Wills Physics Laboratory, University of Bristol, Bristol BS8 1TL, United Kingdom}

\author{Ke-Jin Zhou}
\email{kejin.zhou@diamond.ac.uk}
\affiliation{Diamond Light Source, Harwell Campus, Didcot OX11 0DE, United Kingdom}

\begin{abstract}
Resonant inelastic X-ray scattering (RIXS) is a powerful probe of elementary excitations in solids. It is now widely applied to study magnetic excitations. However, its complex cross-section means that RIXS has been more difficult to interpret than inelastic neutron scattering (INS). Here we report high-resolution RIXS measurements of magnetic excitations of La$_2$CuO$_4$, the antiferromagnetic parent of one system of high-temperature superconductors.  At high energies ($\sim 2$\;eV), the RIXS spectra show angular-dependent $dd$ orbital excitations which are found to be in good agreement with single-site multiplet calculations.  At lower energies ($\lesssim 0.3$\;eV), we show that the wavevector-dependent RIXS intensities are proportional to the product of the single-ion spin-flip cross section and the dynamical susceptibility $\chi^{\prime\prime}(\mathbf{q}, \omega)$  of the spin-wave excitations. When the spin-flip cross-section is dividing out, the RIXS magnon intensities show a remarkable resemblance to INS data. Our results show that RIXS is a quantitative probe the dynamical spin susceptibility in cuprate and therefore should be used for quantitative investigation of other correlated electron materials.     

\end{abstract}

\maketitle

\section{Introduction}
Unconventional superconductors constitute an important group of strongly-correlated electron materials that include heavy fermions, cuprates, ruthenates, and iron-based superconductors.
\cite{stewart1984,bednorz1986,maeno1994,kamihara2006}. They also show a proximity to magnetic ordering or have strong magnetic fluctuations \cite{scalapino2012}. In some cases, the superconductivity can be established by chemical doping or pressurizing a magnetic parent compound.  Although the long-range magnetic order is suppressed by the external tuning, the short-range magnetic fluctuations are found to survive in the superconducting phase of many unconventional superconductors \cite{lyons1988,hayden1996,dai2001,stock2005,lipscombe2007,Headings2011a, dean2015}. The importance of these magnetic fluctuations to the superconducting pairing mechanism has been a subject of many studies in the last decades \cite{scalapino1995,scalapino2012}.       

Experimentally, inelastic neutron scattering (INS) is a well-established probe for the magnetic fluctuations in magnetic materials providing direct measurement of the magnetic structure factor $S(\mathbf{Q}, \omega)$ \cite{Squires1978}. For example, in La$_2$CuO$_4$ (LCO), the parent compound of the first reported cuprate superconductors, INS measurements have observed spin-wave excitations throughout the Brillouin  zone \cite{coldea2001,headings2010}. The excitations are described by large superexchange couplings within the CuO$_2$ planes which extend beyond nearest neighbours. INS has also shown that magnetic excitations persist over a large range wavevectors in superconducting cuprates \cite{Headings2011a}. They are particularly strong near the (1/2,1/2) posistion.  

Compared to INS, RIXS is a newly emerged technique which has been proven to be a powerful tool for probing magnetic excitations in transition metal oxides \cite{ament2009,ament2011resonant}. Owing to its high cross-section and the micron-size focused X-ray beam, RIXS is advantageous over INS in measuring small samples and nanometer-thick films. By working at a resonance, RIXS is element specific thus particularly suited for probing magnetic fluctuations in complex systems with multiple magnetic elements. High-resolution RIXS measurements at the Cu $L$-edge in LCO showed spin excitations with similar dispersion as INS \cite{braicovich2010} and subsequent measurements on a range of doped cuprates have revealed new information about the damping and energy dependence of the spin fluctuations \cite{dean2012,dean2013,dean2013b,guarise2014,wakimoto2015,dean2015,meyers2017}. INS measurements of collective magnetic excitations become technically challenging above about 500\;meV because the background due to multiple scattering becomes large.  RIXS does not suffer from this problem thus it has a major advantage for measuring high-energy excitations. However, the exact RIXS cross-section remains a challenge for fully quantitative interpretation of magnetic excitations. As such, relative comparison is often adopted for tracking intensities evolution as a function of temperature or the doping level \cite{dean2013,dean2013b,le2011intense}. In a previous study of  doped La$_{2-x}$Sr$_x$CuO$_4$, a new procedure was used to determine the absolute wavevector-dependent susceptibility in which magnetic excitation intensities in LCO measured by RIXS were assumed to be equivalent to those obtained from INS \cite{robarts2019anisotropic}.

Here we aim to improve the understanding of RIXS as a probe of magnetic excitations by characterising its cross-section. LCO is chosen because of extensive studies on its magnetic excitations made by INS. The RIXS measurements were performed on LCO single crystals with surface normals (001) and (100) which provide a good test for the geometrical dependent single ion cross-section of the orbital and the spin-flip excitations. The sample with the surface normal (100) provides further access along the $(h, h)$ direction in reciprocal space compared to previous RIXS measurements. We firstly compare the measured $dd$ orbital excitations and the calculations using crystal-field theory (CFT) implemented in the many-body code QUANTY \cite{haverkort2012multiplet,haverkort2016quanty}. Good agreement was obtained for both samples. The comparison was then extended to the magnetic excitations. The extracted wavevector-dependent suscepitibilities $\chi^{\prime}(\mathbf{Q})$ show a remarkable match with those seen by INS after dividing out the spin-flip excitations. We highlight that the established method can be readily applied to quantitative studies of magnetic excitations in other superconductors, such as ruthenates, nicklelates, as well as general transition-metal oxides based magnetic materials.

\section{RIXS experiments}
The RIXS experiments were performed on single crystal samples of LCO which were grown using the travelling solvent floating zone technique. We describe LCO using its high-temperature tetragonal (HTT) I4/mmm crystal structure in which $a$ = $b \simeq$ 3.8\;\AA, $c \simeq$ 13.2\;\AA. The momentum transfer $\mathbf{Q}$ is defined in reciprocal lattice units (r.l.u.) as $\mathbf{Q} = h\mathbf{a^{*}} + k\mathbf{b^{*}} + l\mathbf{c^{*}}$ where $\mathbf{a^{*}} = 2\pi/a$ etc. The energy of the scattered photons is given by $\hbar \omega =c\left|\mathbf{k}\right| - c\left|\mathbf{k}^{\prime}\right|$ and momenta $\mathbf{Q} = \mathbf{k} - \mathbf{k}^{\prime}$, where  $\mathbf{k}$ and $\mathbf{k}^{\prime}$ is the incident and scattered photon wavevector, respectively.

\begin{figure}
\centering
\includegraphics[width=\linewidth]{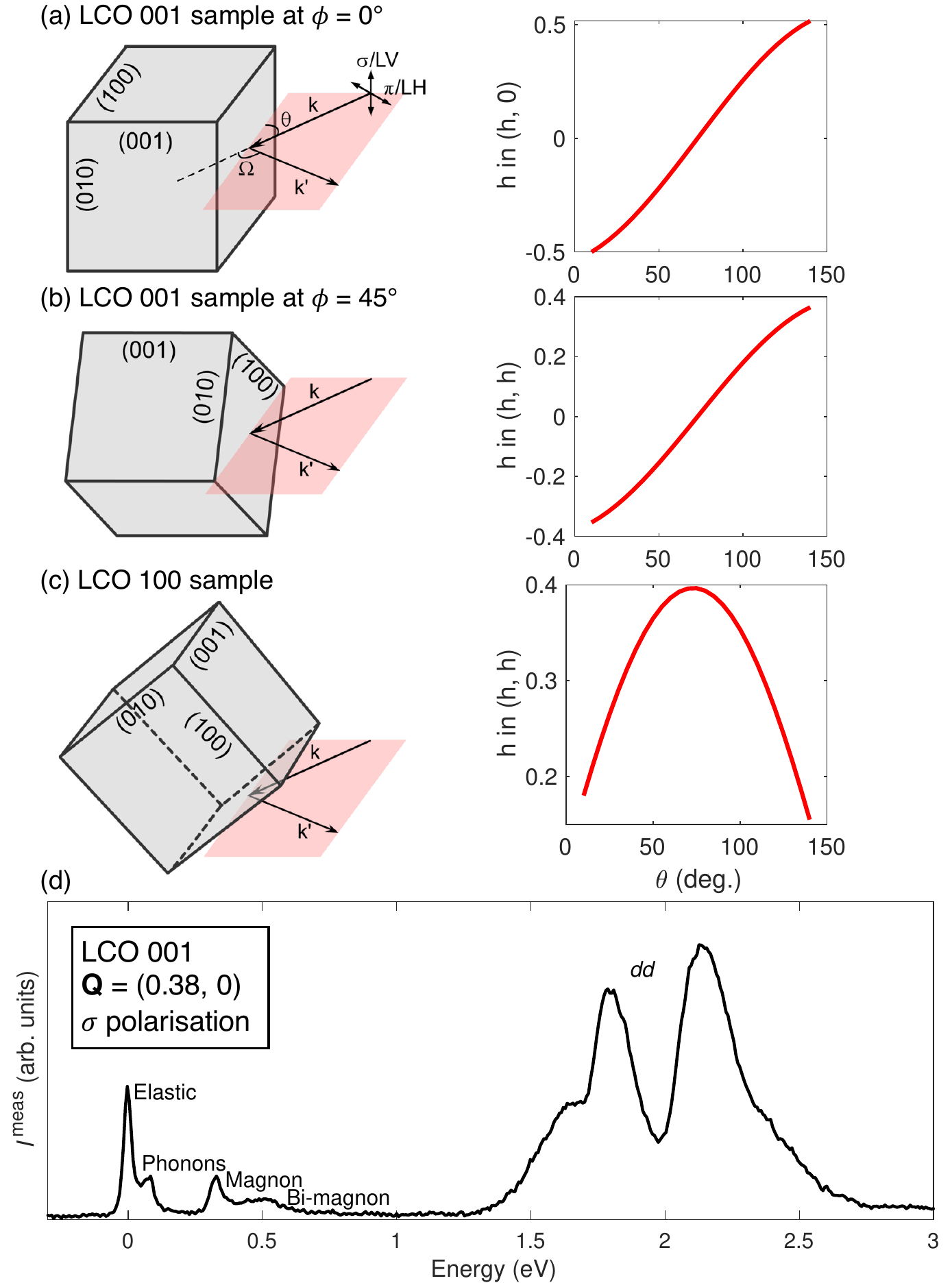}
\caption{RIXS experimental geometries for the two samples. (a) The LCO001 sample at $\phi = 0^{\mathrm{o}}$, probing $(h, 0)$. (b) The LCO001 sample at $\phi = 45^{\mathrm{o}}$, probing $(h, h)$. (c) The LCO100 sample probing $(h, h)$. For each orientation, the projection of the momentum transfer and the relationship between $\theta$ and the momentum transfer is shown. Panel (d) shows various excitations resolved in a typical RIXS spectrum.}
\label{fig:lco_orientation}
\end{figure}

High-resolution RIXS spectra were acquired at the I21-RIXS Beamline at Diamond Light Source, United Kingdom. The incoming X-ray photon energy was tuned to the Cu $L_3$ resonance ($\simeq$ 931.5\;eV) and we performed measurements with both linear horizontal (LH)/ $\pi$ polarisation and linear vertical (LV)/ $\sigma$ polarisation (Fig. \ref{fig:lco_orientation}). The total instrumental energy resolution has standard Gaussian distribution with the full width at the half-maximum of $\Delta E \simeq$\;37\;meV. All RIXS measurements were conducted at 15 K. 

Two different samples, LCO001 and LCO100, were prepared with the surface normals approximately (001) and (100) respectively. The samples were aligned and cleaved in-situ to expose a clean surface to the beam. Fig. \ref{fig:lco_orientation} (a, b) shows how data were collected from LCO001 along $(h,0)$ and $(h,h)$ respectively.  Fig.~\ref{fig:lco_orientation}(c) shows how LCO100 was mounted on a 45$^{\circ}$ wedge such that $(1\bar{1}0)$ was perpendicular to the scattering plane. This allowed data to be collected along $(h,h)$ to larger $h$. Because of the two-dimensionality of the magnetic excitations in cuprates, the in-plane wavevector $(h,k)$ is varied by scanning the angle of incidence $\theta$ while keeping the scattering angle, $\Omega$, fixed at 154$^{\circ}$. The zero of $\theta$ ($\theta=0$) is defined such that (110) is anti-parallel to $\mathbf{k}$ and ``grazing-in'' $\mathbf{k}$ probes negative $h$.  Thus with LCO001, we access $(h, 0)$ = (-0.5, 0) to (0.5, 0) and $(h, h)$ = (-0.35, -0.35) to (0.35, 0.35). For LCO100, $(h, h)$ is probed with a maximal in-plane wavevector of (0.4,0.4).

A typical RIXS spectrum is shown in Fig. \ref{fig:lco_orientation} (c) in which a quasi-elastic peak, phonon excitations, single-magnon, multimagnon, and $dd$ orbital excitations are clearly resolved. The measured RIXS intensity, $I_{\sigma(\pi)}^{\mathrm{meas}}$ can be corrected to yield the real RIXS intensity $I_{\sigma(\pi)}^{\mathrm{corr}}$, to account for energy, wavevector and polarisation-dependent self-absorption effects. The self-absorption correction method is an extension of the simple procedure in recent works \cite{minola2015,kang2019resolving} and is described explicitly in Appendix \ref{apx_ss:SA}. In Section III, we will firstly focus on the $dd$ orbital excitations and in Section IV, we will discuss the magnetic excitations. 

\section{$dd$ orbital excitations}

\begin{figure}
\centering
\includegraphics[width=\linewidth]{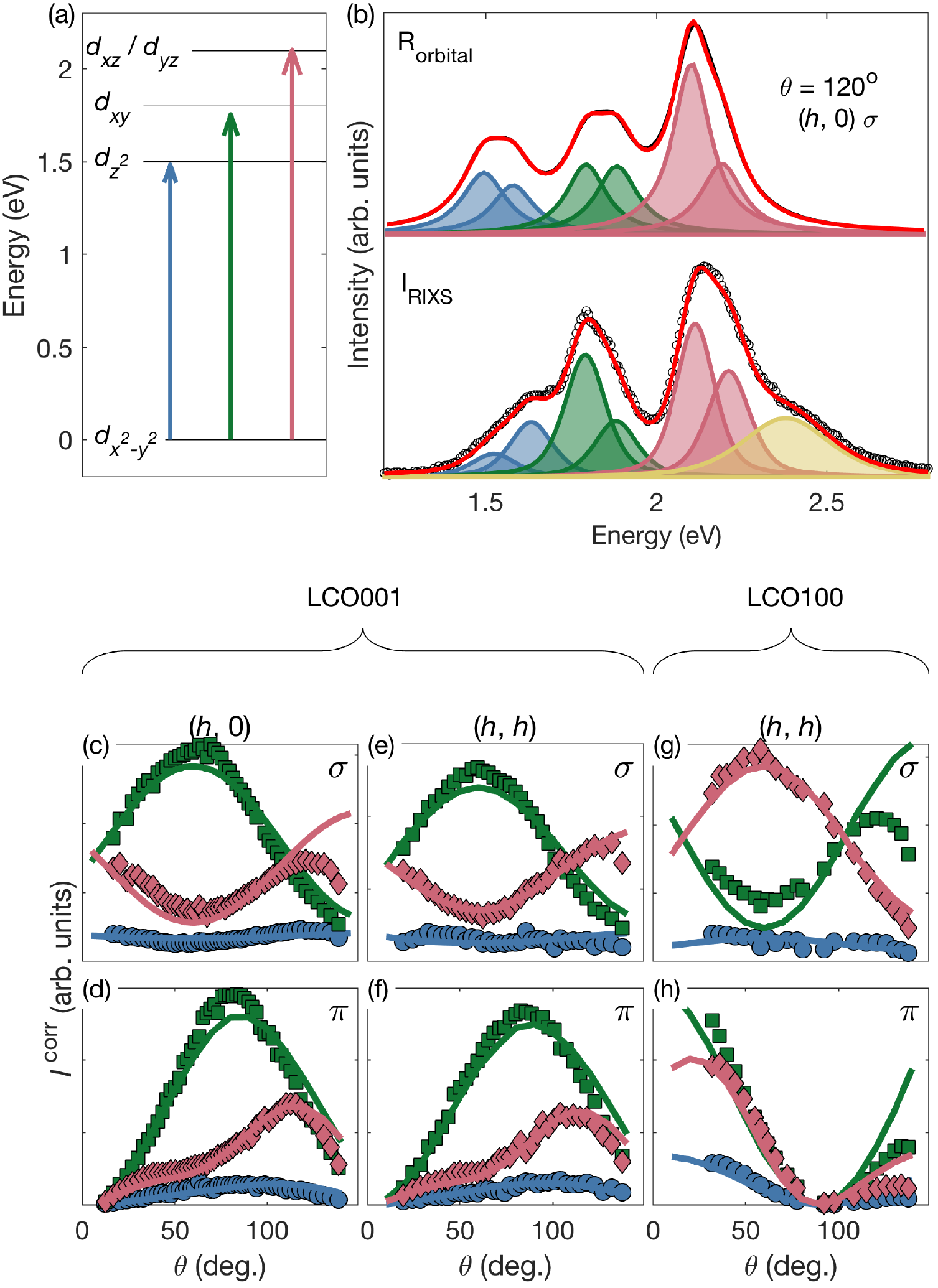}
\caption{Orbital excitations in RIXS showing (a) the $dd$ transitions in LCO and (b) a comparison of the calculated $dd$ peaks from crystal field theory (CFT) and RIXS measurements for a typical spectra at $\theta$ = 120$^{\circ}$ along the $(h, 0)$ direction, LCO001 orientation and with $\sigma$ polarisation. Panels (c-h) show the intensity of the $dd$ excitations from fitted RIXS data compared with the relative intensities calculated in CFT. }
\label{fig:compare_orbitals_pi0_LV}
\end{figure}

The RIXS spectra are dominated by strong $dd$ excitations between 1.3 and 3\;eV which occur due to transitions between the ground  and the excited 3$d$ orbital states as sketched in Fig. \ref{fig:compare_orbitals_pi0_LV} (a). $dd$ excitations have been extensively studied using RIXS in the past for understanding the local crystal field splittings among various cuprate families \cite{ghiringhelli2004,moretti2011}. Plotted in Fig. \ref{fig:compare_orbitals_pi0_LV} (b) is a representative RIXS spectrum in the energy range of the $dd$ excitations. A pseudo-Voigt function is used for the spectral fitting where each $dd$ excitation seems to comprise two peaks. The higher energy peak in each orbital excitation is due to the mixture with the spin-flip. Such excitation has not been observed in previous experiments due to the limitation of the energy resolution \cite{moretti2011}. There is also additional spectral weight present at a higher energy of 2.4\;eV. This peak is seen in a previous work and has been attributed to oxygen vacancies which are thought to alter the crystal field acting on the Cu ions \cite{moretti2011}. We fit this peak with an additional pseudo-Voigt function.

To reproduce the experimental observations, we used the single-site crystal-field multiplet theory (CFT) implemented in the many-body QUANTY code \cite{haverkort2012multiplet,haverkort2016quanty} to calculate the $dd$ excitations. To simulate the spin-flip excitation, an inter-atomic exchange integral of 100\;meV is added along the $(h, h)$ direction of the CuO$_2$ planes. Throughout the paper, all calculations were done with the outgoing polarisations effect averaged. Appendix \ref{apx:CFT} describes the calculation details. We assume that in the range of the $dd$ excitations, the self-absorption corrected RIXS intensity can be described as a product of a pre-factor $f'$ and the single-ion orbital cross-section $R_{\mathrm{orbital}}(\boldsymbol{\epsilon},\boldsymbol{\epsilon}^{\prime}, \mathbf{k}, \mathbf{k}^{\prime})$:
\begin{equation}
   I_{\mathrm{orbital}} = f'\times R_{\mathrm{orbital}}(\boldsymbol{\epsilon},\boldsymbol{\epsilon}^{\prime}, \mathbf{k}, \mathbf{k}^{\prime}).
\end{equation}
Fig. \ref{fig:compare_orbitals_pi0_LV} (b) compares the intensity of the orbital component, $R_{\mathrm{orbital}}$, to the RIXS measurements for a representative spectra. The calculations seem to agree well with the experiments.

$dd$ orbital excitations possess a strong angular dependence owing to the anisotropy of the Cu 3$d$ orbitals \cite{moretti2011}. In Fig. \ref{fig:compare_orbitals_pi0_LV} (c-h) we summarize the integrated $dd$ orbital excitation intensity as a function of $\theta$ under various experimental configurations. Note that each data point represents the sum of the integrated area of normal $dd$ and the spin-flip assisted $dd$ excitations.

Theoretically, $\theta$-dependent $dd$ excitations in LCO have been simulated successfully using the single-ion model \cite{ament2009, moretti2011}. Here we use QUANTY to perform similar calculations. Fitted area of each calculated $dd$ excitation is superimposed on the experiment data shown in Fig. \ref{fig:compare_orbitals_pi0_LV} (c-h). Note that the calculated $dd$ excitation intensities have been independently scaled to their respective orbitals. The overall comparison between the experiment and the theory yields a good consistency except some deviations at small grazing out angles. For instance, the d$_{xz}$/d$_{yz}$ orbital in LCO001 with $\sigma$ polarization and the d$_{xy}$ orbital in LCO100 with both polarizations. This may be due to the accuracy of the fit at those extreme angles. The good agreement of the $\theta$-dependent $dd$ intensities suggest that the local $dd$ cross-sections based on a single site multiplet theory reproduce the experimental data from samples with different crystal orientations very well. We are therefore highly motivated to apply the theory to the magnetic excitations whose intensities are strongly $\theta$ dependent  \cite{ament2009} . 

\begin{figure*}
\centering
\includegraphics[width=\linewidth]{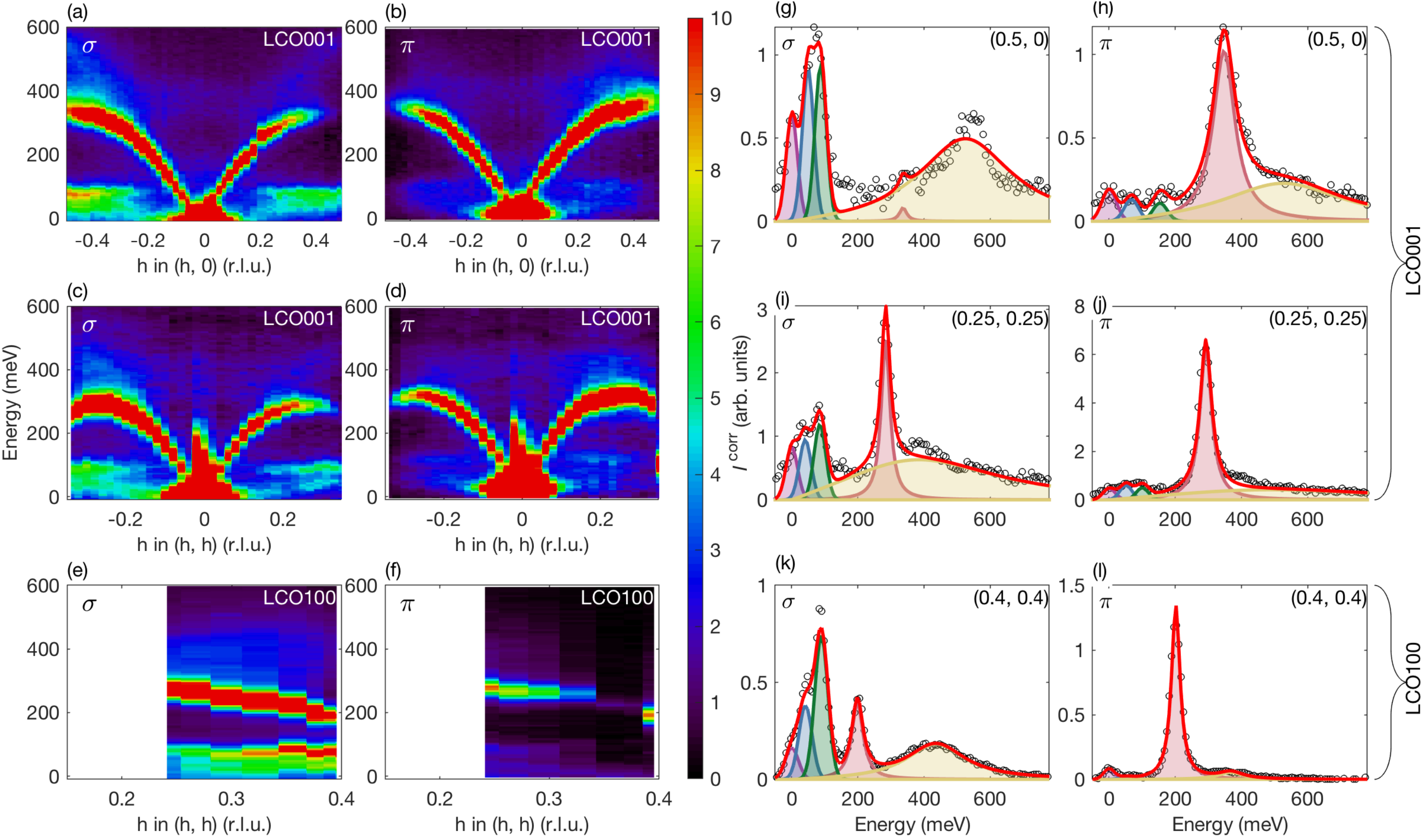}
\caption{The low-energy excitations in LCO showing (a-f) self-absorption corrected RIXS intensity maps and (g-l) representative RIXS spectra. Total fits to the data are shown in red. The DHO magnon and multimagnon fit are shown in pink and yellow respectively. Gaussian fits to the elastic peak, phonon and multiphonon peaks are in purple, blue and green respectively.}
\label{fig:intensity_map_examples}
\end{figure*}

\section{Dynamical spin susceptibility} 
Fig. \ref{fig:intensity_map_examples} (a-f) show RIXS intensity maps in the low energy region as a function of the momentum transfer along (100) and (110) directions. Near zero energy loss, we see quasi-elastic peaks. Between zero and 100 meV, two phonon branches are clearly resolved throughout the accessible momentum space. These are likely the bond-buckling and the bond-stretching modes \cite{devereaux2016directly} which will be discussed in a separate work. As is known, the strongly dispersive features are single magnons emanating from the zone center to the zone boundary up to almost 400\;meV \cite{coldea2001, headings2010}. In particular, single magnons in LCO100 show consistent dispersion compared to that in LCO001 projected along the $(h, h)$ direction. Beyond the zone boundary of $(1/4, 1/4)$, the single magnons continue to disperse to lower energy akin to the INS data (Fig.3(c)-(f)) \cite{coldea2001, headings2010}. Broader peaks appear between 400\;meV and 600\;meV which are most likely the multimagnons as observed by RIXS at the Cu L$_3$- and the O K- edges \cite{bisogni2012,chaix2018resonant}.

The spectra between --80 and 800\;meV are modelled with Gaussian functions to account for the elastic peak and phonons and with the response function of a damped harmonic oscillator (DHO) to account for the (single and multi) magnon excitations. The DHO model has been used in several RIXS studies \cite{monney2016,peng2018,robarts2019anisotropic} and describes the response for a range of damping. We fit the spectra to the imaginary part of the DHO response, given by,
\begin{equation}
\label{eqn:damping}
\chi^{\prime\prime}(\mathbf{Q}, \omega) = \frac{\chi^{\prime}(\textbf{Q}) \, \omega_0^2(\mathbf{Q)} \, \gamma(\mathbf{Q)}\, \omega}{\left[\omega^2 -\omega_0^2(\mathbf{Q)}\right]^2+\omega^2\gamma^2(\mathbf{Q)}},
\end{equation}
where $\chi(\mathbf{Q})$ is the real part of the susceptibility at zero frequency, $\omega_0$ describes the position of the excitation pole and $\gamma$ represents the damping.

\begin{figure}
\centering
\includegraphics[width=\linewidth]{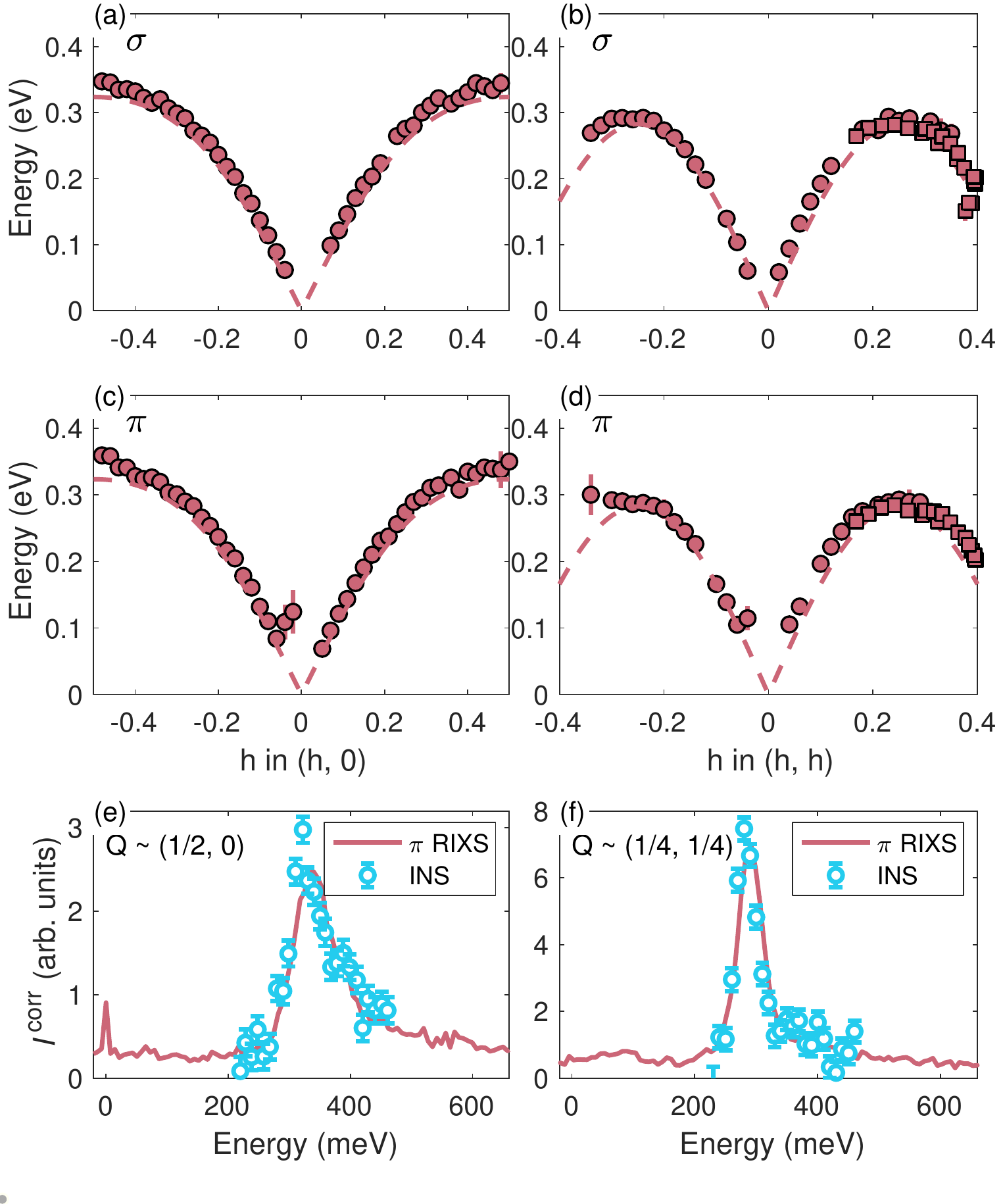}
\caption{(a-d) show the energy dispersion of magnon excitations measured by RIXS. Red symbols indicate $\omega_0$ extracted from the damped harmonic oscillator fit to the magnons and data from the LCO001 and LCO100 orientations are indicated with circles and squares respectively. The dashed line shows the magnon dispersion extracted from INS \cite{headings2010}. Panels (e-f) show comparison of the magnon lineshape measured with $\pi$ polarised RIXS in pink and INS in cyan, at wavevectors close to $(1/2, 0)$ and $(1/4, 1/4)$. The INS data are scaled to the RIXS data and a constant offset of 0.5 is added to the INS to compensate for a possible over subtraction of the background in Ref. \cite{headings2010}.}
\label{fig:all_excitations_energies}
\end{figure} 

Examples of the RIXS spectra and fittings are shown in Fig. \ref{fig:intensity_map_examples} (g-l). The data show drastically different ratio between the single and multimagnon intensity measured with incident polarisation $\sigma$ or $\pi$. At \textbf{Q} = $(1/2, 0)$, the single magnon component dominates the spectra for $\pi$ polarisation whilst it becomes much weaker and almost entirely obscured by the multimagnon component with $\sigma$ polarisation. The drastic polarisation and $\theta$ dependence is due to the local spin-flip cross-section \cite{ament2009}. Strong multimagnon scattering is observed at the enigmatic region $(1/2, 0)$ under the $\sigma$ polarisation where the profile are distinct to those in the rest of the reciprocal space.

To illustrate the single magnon dispersion more clearly, we plot the magnetic pole $\omega_0$ of LCO001 in Fig. \ref{fig:all_excitations_energies} (a-d). On top of that, we add data points obtained from LCO100 along the $(h, h)$ direction. Noticeably, the single magnon dispersion obtained from LCO001 and LCO100 matches very well. To compare with the spin-wave theory (SWT), we computed the dispersion using the nearest and the next-nearest neighbour exchange constants extracted from INS \cite{headings2010}. Good agreement is seen in both $(h, 0)$ and $(h, h)$ direction. Fig. \ref{fig:all_excitations_energies} (e) and (f) shows comparisons of the magnon spectra obtained between RIXS and INS at two zone boundary positions. Remarkably, the lineshape of the magnon spectra agree well between two techniques. At the $(1/4, 1/4)$, the magnon excitation shows a resolution-limited peak whereas at $(1/2, 0)$, both RIXS and INS spectra present some spectral weight at high energy as a continuum.

\begin{figure*}
\centering
\includegraphics[width=\linewidth]{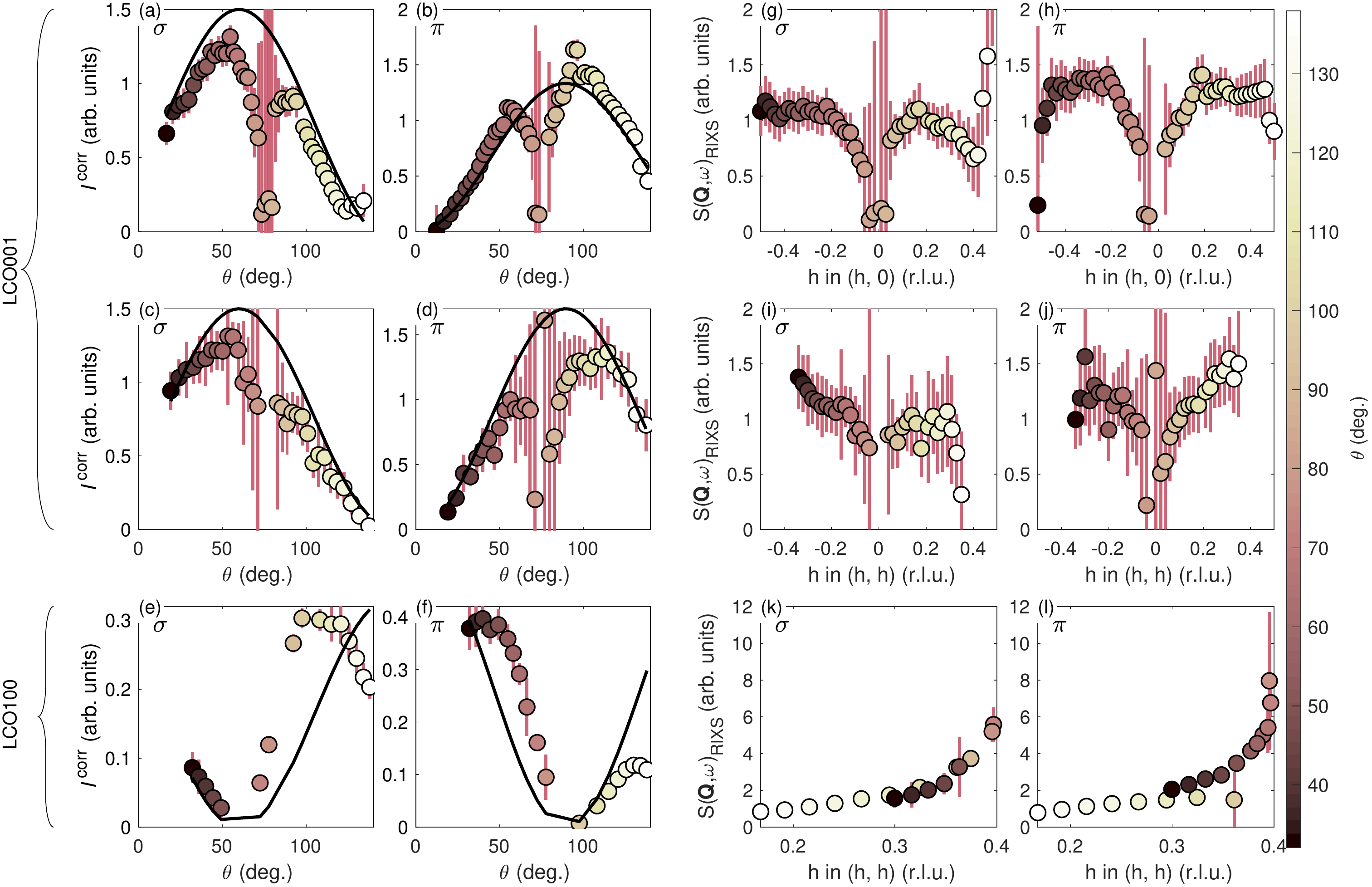}
\caption{Details of the deconvolution procedure for the magnon intensity. Panels (a-f) show the self-absorption corrected magnon intensity, $I_{\mathrm{corr}}$ (filled circles), compared to the single-ion spin-flip cross-section $R_{\mathrm{spin}}$ (black line). Panels (g-l) show the fully deconvoluted intensity, $S(\mathbf{Q},\omega)_{\mathrm{RIXS}}$ as a function of \textbf{Q}. The value of $\theta$ is indicated by the colour.}
\label{fig:magnon_intensity_corrected}
\end{figure*}

We now discuss the RIXS intensity of the single magnons. As with the analysis of the $dd$ excitations, the single magnon intensities are extracted from an integration of the DHO function and summarized in Fig. \ref{fig:magnon_intensity_corrected} (a-f). Note that these data are presented as a function of $\theta$ due to the complex projection along the $(h, h)$ direction in LCO100. Importantly, as the single magnons are strongly dispersive in the energy-momentum space, an accurate self-absorption correction is performed for both the energy and momentum dependence. Details of the self-absorption correction are presented in Appendix \ref{apx_ss:SA}.
It is generally accepted that under certain conditions \cite{ament2011resonant}, the RIXS intensity is proportional to the dynamic structure factor $S(\mathbf{Q},\omega)$ multiplied by a resonant factor $f$ which is dependent on the polarisation $\boldsymbol{\epsilon}$ and $\boldsymbol{\epsilon}^{\prime}$ of the initial and final photons with wavevector  $\mathbf{k}$ and $\mathbf{k}^{\prime}$. As the single-ion spin-flip cross-section is controlled by the RIXS process, we further express the resonant factor $f(\boldsymbol{\epsilon},\boldsymbol{\epsilon}^{\prime}, \mathbf{k}, \mathbf{k}^{\prime})$ as a product of a pre-factor $f'$ and the single-ion spin-flip cross-section $R_{\mathrm{spin}}(\boldsymbol{\epsilon},\boldsymbol{\epsilon}^{\prime}, \mathbf{k}, \mathbf{k}^{\prime})$: 
\begin{equation}
  I_{\mathrm{spin}} = f'\times R_{\mathrm{spin}}(\boldsymbol{\epsilon},\boldsymbol{\epsilon}^{\prime}, \mathbf{k}, \mathbf{k}^{\prime}) \times S(\mathbf{Q},\omega).
\end{equation}  
Under this approximation, we computed the single-ion spin-flip cross-section, $R_{\mathrm{spin}}$, as a function of $\theta$ using the parameters optimised for the $dd$ calculations. The calculated spin-flip spectra were fitted using DHO function (Eqn. \ref{eqn:damping}) and the integrated spectral weight are plotted on top of the experimental data in Fig. \ref{fig:magnon_intensity_corrected} (a-f).  

The first glimpse informs us that the experimental magnon intensities modulate as a function of $\theta$ and follow the overall trend of the local spin-flip cross-section at least for LCO001. However there are deviations around $\theta\simeq$75$^{\circ}$ where the magnon intensity drops almost to zero. This is probably due to the diminished spin wave intensity near the zone center. It is thus appealing to see both the local and the collective behaviors play a role in the magnon intensities in RIXS. For LCO100, it is less obvious to trace the evolution of the magnon intensities between the experiment data and the local spin-flip cross-section. This may be due to the unusual projection of the momentum transfer along the $(h, h)$ direction. 

To reveal the dynamic spin susceptibility from RIXS, we  divided the magnon intensity by the local RIXS spin cross-section from the CFT calculations,
\begin{equation}
    {S(\mathbf{Q},\omega)}_{\mathrm{RIXS}} \propto \frac{I_{\mathrm{spin}}}{R_{\mathrm{spin}}}.
\end{equation}
${S(\mathbf{Q},\omega)}_{\mathrm{RIXS}}$ is shown for both LCO001 and LCO100 in Fig. \ref{fig:magnon_intensity_corrected} (g-l) as a function of wavevector. Remarkably, the simple process yields a highly symmetrical intensity profile with respect to the zone center in LCO001 along both $(h, 0)$ and $(h, h)$ directions. The symmetrical magnon intensity profile is reminiscent of the spin-wave intensity in the Heisenberg model \cite{coldea2001, headings2010}. For LCO100, the difference after the removal of the local spin-flip cross-section is even more striking. The irregular magnon intensity profile evolved to a clear exponential-like trend as a function of \textbf{Q} along the $(h, h)$ direction. We want to bring the attention to the effect of the incident X-ray polarizations. The fact that there is almost no polarization dependence among all sets of data pointing to a simple message, that is, the pure collective dynamic spin susceptibility is independent of the property of the experimental probe. It is also worth mentioning the error bars of the magnon intensities. For LCO001, data points near the zone center are marked with large error bars due to the uncertainty of fitting the negligible raw magnon intensities. Similarly, the data points near the very grazing-out (grazing-in) geometry in $\sigma$($\pi$) polarization are associated with large error bars as a result of minimal local spin-flip cross-section. The latter reason also applies to LCO100 where the large error bars are present near $\theta\simeq$ 60$^{\circ}$ and 90$^{\circ}$ for $\sigma$ and $\pi$ polarizations, respectively.

\begin{figure}
\centering
\includegraphics[width=\linewidth]{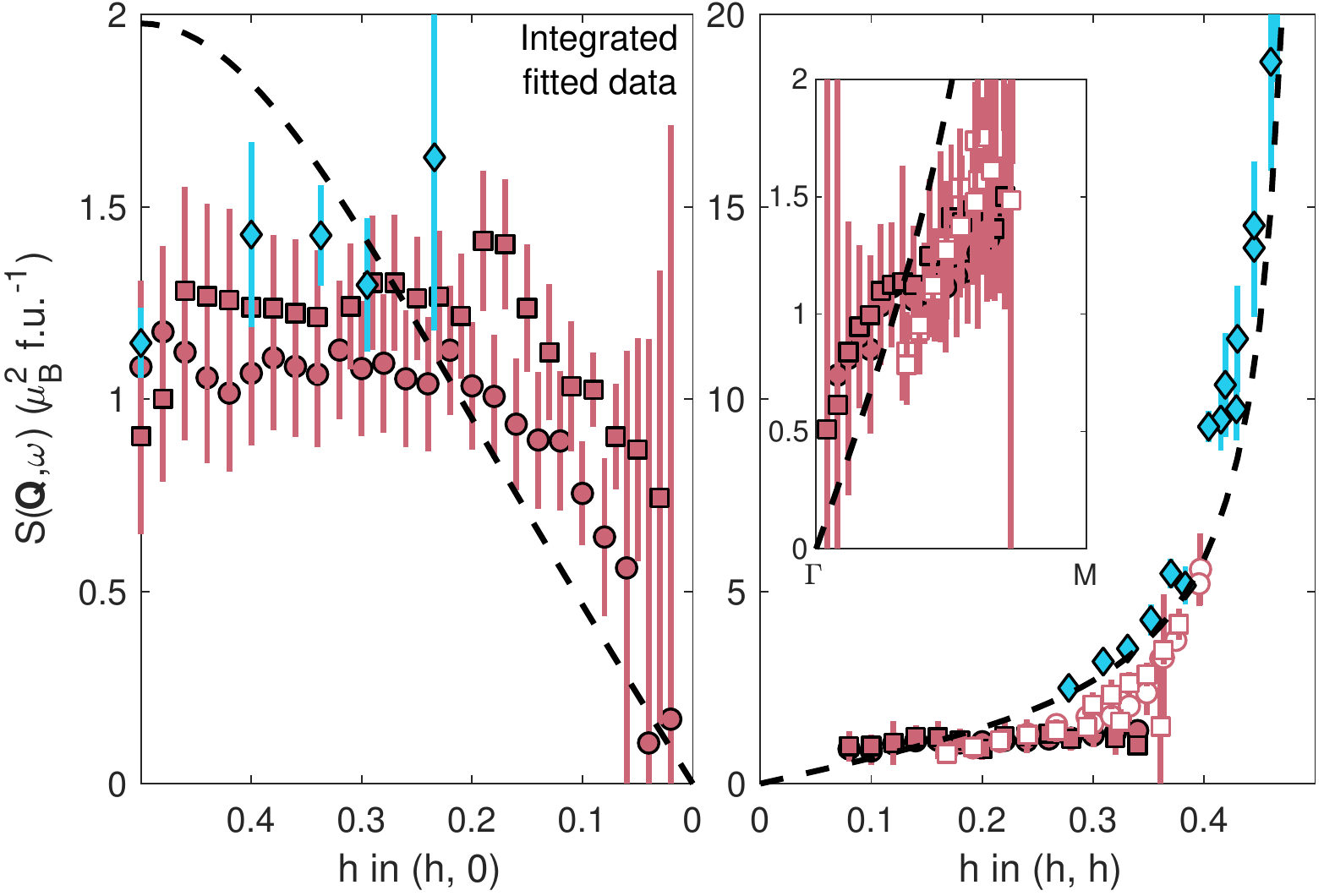}
\caption{Comparison of the spin-wave intensity for RIXS, INS and the theoretical values from SWT. INS data are shown as blue diamonds and the SWT as a black dashed line. RIXS data from the LCO100 and LCO100 orientation are indicated in pink and white marks, respectively, and $\sigma$ and $\pi$ polarisation are represented by circles and squares, respectively. }
\label{fig:magnon_rixs_ins}
\end{figure}

Fig. \ref{fig:magnon_rixs_ins} shows $S(\mathbf{Q},\omega)_{\textrm{RIXS}}$ where regions with relative small error bars are selected, i.e., $\sigma$ grazing-in, $\pi$ grazing-out for LCO001, while $\sigma$ grazing-out and $\pi$ grazing-in for LCO100. We also plot the magnon intensity measured with INS by Headings \textit{et al.} \cite{headings2010} for the overlapped momentum space. Note that a single scaling constant is applied to the RIXS data in order to compare with INS data. In addition, we plot the spin-wave intensity calculated from the next-next nearest neighbour SWT model.

Along the $(h, 0)$ direction, both INS and RIXS show an approximately constant magnon intensity profile from 0.2 to 0.5 $r.l.u.$. Such behavior is a clear deviation from the linear spin-wave theory which increases monotonically approaching the zone boundary. The effect is explained in INS study due to the creation of a spinon pair \cite{headings2010}. INS measurements were only made up to 450 meV and it is interesting to note that RIXS measurements up ~1\;eV yield similar results. This in turn demonstrates that the discrepancy at the zone boundary of $(1/2, 0)$ between the experimental results and the SWT is genuine. Along the $(h, h)$ direction, LCO100 shows a sharp intensity increase towards the \textbf{Q}$_{\mathrm{AFM}}$ = (1/2, 1/2). We see vague indication of the trend in LCO001, however, the measurements do not reach large enough \textbf{Q}. The measurements on LCO100 allows us to reach $(0.4, 0.4)$ which is crucial for the comparison with INS experimental data. The level of agreement is good among RIXS, INS and SWT model which all show an increased intensity toward the \textbf{Q}$_{\mathrm{AFM}}$ = (1/2, 1/2). At low \textbf{Q}, RIXS generally observes greater spectral weight than INS. The additional intensity may well result from the specular reflection influencing the magnon fits. 

The intensity of the multimagnon excitations, which are well resolved in our data, could be extracted and corrected in the same way as the single magnons. However, in this instance the scattered polarisation is not well understood and the photon out polarisation analysis is necessary to separate components of the multimagnon continuum. The single-site calculations are also not adequate in this case as multimagnon creation is likely to be more complex  \cite{ament2009,ament2010strong}. Cluster calculations are needed to provide a proper description of the multiple-sites spin-flip cross-section before the extraction of the dynamical spin susceptibility of the multimagnons.

\section{Conclusions and outlook}
We have made high-resolution RIXS measurements of the orbital and collective magnetic excitation for La$_2$CuO$_4$ single crystals with the surface normals (001) and (100). The spin-flip assisted $dd$ orbital excitations are clearly resolved owing to the high energy resolution. Strong $\theta$-dependence of the orbital excitations are well reproduced by the multiplet crystal field theory. The momentum-dependent collective magnetic excitations are measured along both $(h, 0)$ and $(h, h)$ directions of the first Brillouin zone to the extent that is possible at the Cu L$_3$-edge. The dispersion of the single magnons show a very good match to the spin-wave theory. Remarkably, the RIXS single magnon spectral profiles are reminiscent of those measured by INS. We determined the wavevector-dependent single-magnon response by correcting for the self-absorption and the local spin-flip RIXS matrix element. It is found that this response reflects the known dynamical spin susceptibility for La$_2$CuO$_4$ regardless of the incident photon polarization and the many-body effects involved in the RIXS process. Comparing to INS data, RIXS show excellent agreement along both primary directions. In particular, the consistent results between RIXS and INS show strong deviation from the spin-wave theory indicating the abnormal spin susceptibility approaching $(h, 0)$ zone boundary is a genuine behavior of the system.       

Our results have several important implications. The established method of retrieving the pure spin susceptibility is readily applicable to doped cuprate superconductors. This is due to the existence of the fundamental matrix elements irrespective of the parent or the doped cuprate compounds. RIXS is able to make accurate reliable measurements of the spin susceptibility. Furthermore, the new method is also beneficial to spin susceptibility studies of many magnetic transition metal oxides. For instance, RIXS has revealed the magnetic excitations in 214 nickelates whose dispersions are consistent to that obtained with INS \cite{fabbris2017doping,nakajima1993spin}. As the local spin-flip cross-section can be easily computed in these 3d$^8$ systems, the study of the spin susceptibility will enable closer comparison with INS results. Finally, RIXS is unique in probing the higher-order, such as quadrupolar, magnetic excitations \cite{nag2020many}. Applying the method to the study of the spin susceptibility of these higher-ranked magnetic orders is vital to systems like Kitaev quantum spin liquids.  

\acknowledgements
         K.J.Z. and S.M.H. acknowledge J. van den Brink for fruitful discussions. The authors acknowledge funding and support from the Engineering and Physical Sciences Research Council (EPSRC) Centre for Doctoral Training in Condensed Matter Physics (CDT-CMP), Grant No. EP/L015544/1 as well as Grant No. EP/R011141/1. We acknowledge Diamond Light Source for providing the beamtime under the proposal SP18469 and the science commissioning time on the Beamline I21. We acknowledge Thomas Rice for the technical support throughout the beamtime. We would also like to thank the Materials Characterisation Laboratory team for help on the Laue instrument in the Materials Characterisation Laboratory at the ISIS Neutron and Muon Source.

\appendix
\section{Experimental details}

\subsection{Sample preparation}
Samples of single-crystal LCO were grown via the travelling-solvent floating zone technique (TSFZ), annealed in an Argon atmosphere to remove excess oxygen, detwinned and cleaved \textit{in-situ}. The crystals were previously used in the neutron scattering measurements described in reference \cite{headings2010}.

\subsection{Data processing}
RIXS data are extracted by integrating along the non energy-dispersive direction at each \textbf{Q} after subtracting the dark-image background. Spectra are normalised by the counting time. The zero-energy positions of RIXS spectra were determined by comparing to reference spectra recorded from the amorphous carbon tapes next to the sample for each \textbf{Q} position. They were finely adjusted through the Gaussian fitting of each elastic peak. It is clear that this process is much easier close to the specular position (\textbf{Q} = 0 in the LCO001 orientation) where the elastic peak becomes large. To reflect this, the error in the energy correction is established by the error in fitting a Gaussian peak multiplied by a Bose function $n(\omega) + 1$ to model the excitations near $\omega = 0$. The shift in energy remains within 12\;
meV throughout the \textbf{Q}-range, therefore we conclude that the procedure is consistent regardless of the intensity of the elastic peak.

\subsection{Self-absorption correction}
\label{apx_ss:SA}
\begin{figure}
\centering
\includegraphics[width=\linewidth]{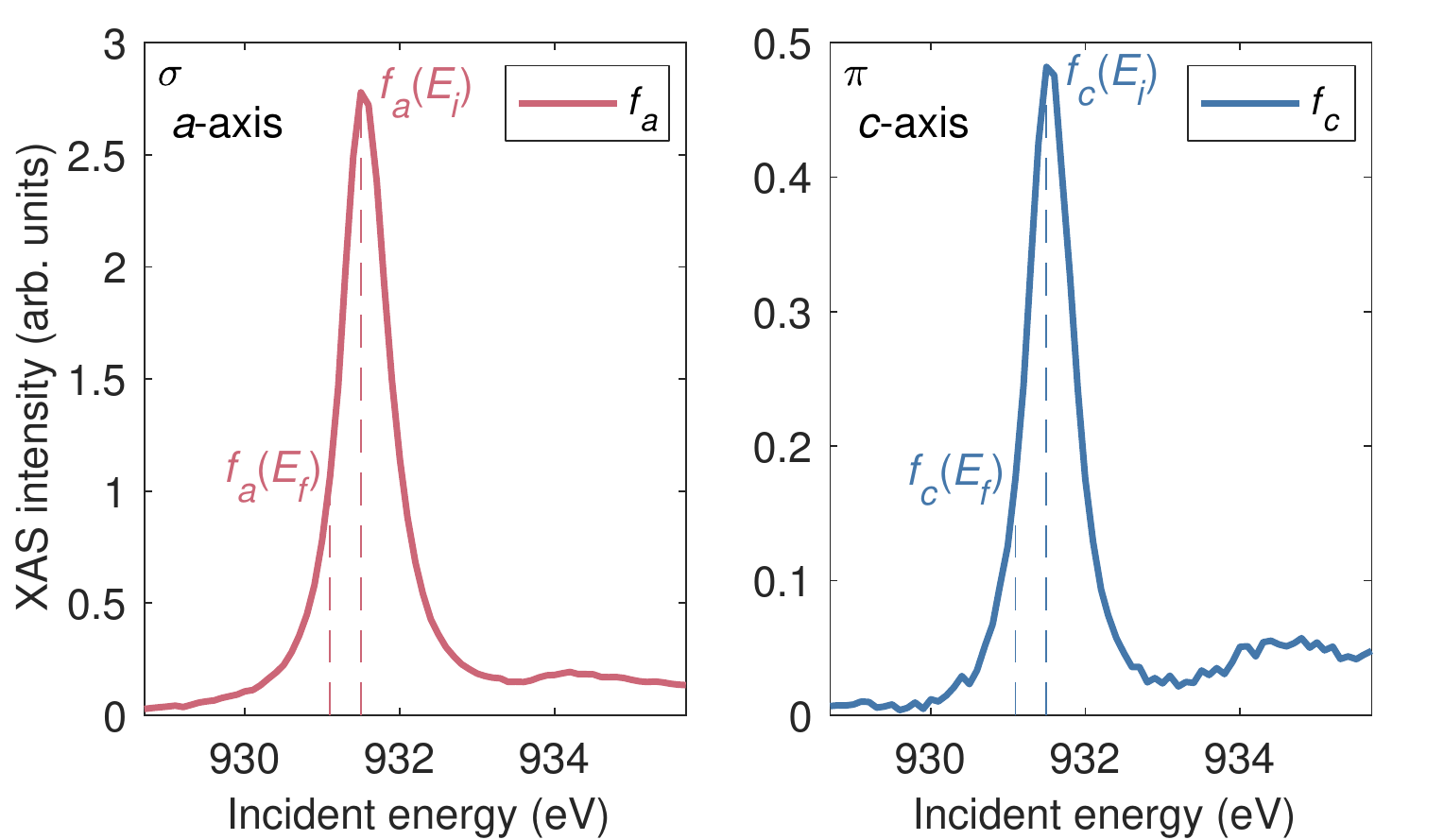}
\caption{The projected XAS spectra along the \textit{a}- and \textit{c}- axes in LCO samples with $\sigma$ and $\pi$ polarisations. $f_{a}(E_i), f_{c}(E_i)$ have fixed intensity due to the fixed incident energy while $f_{a}(E_f),f_{c}(E_f)$ of emitted X-rays are energy dependent.}
\label{fig:compare_XAS}
\end{figure}
We follow recent practice \cite{comin2015, comin2015b, minola2015, achkar2016,fumagalli2019} to correct for the effects of self-absorption in our data using x-ray absorption spectra (XAS) measured in the same geometry as the RIXS measurements. We estimate a self-absorption factor, $C_{\mathrm{SA}}$ using the same procedure as outlined in reference \cite{robarts2019anisotropic}, where the corrected intensity $I_{\sigma(\pi)}^{\mathrm{corr}}$ is  $I_{\sigma(\pi)}^{\mathrm{corr}} = C_{\mathrm{SA}}I_{\sigma(\pi)}^{\mathrm{meas}}$ or,
\begin{equation}
I_{\sigma(\pi)}^{\mathrm{corr}} = I_{\sigma(\pi)}^{\mathrm{meas}}  \frac{\mu_{i,\sigma(\pi)}\sin(\Omega - \theta) + \mu_{f,\sigma(\pi)}\sin\theta}{\sin(\Omega - \theta)}.
\label{eqn:self_absorption}
\end{equation}
Here $\mu_i$ and $\mu_f$ are the absorption coefficients extracted from XAS performed prior to the RIXS measurements. In our experiments the XAS is measured with the total electron yield. Components of the photon form factor, $f_{a}$ and $f_{c}$, are extracted from the XAS intensity when the electric field of the incident x-rays is parallel to the crystalline $a$ and $c$ axes respectively which can be accessed with $\sigma$ and $\pi$ polarised x-rays respectively. Fig. \ref{fig:compare_XAS} shows example XAS spectra measured along the $a$ and $c$-axes. $f_{a}$ and $f_c$ are found from the intensity of the XAS spectra at $E_i$ and $E_f$ relative to the intensity at resonance. These values allow us to estimate the absorption coefficients for the incident and outgoing x-rays in the LCO001 geometry,
\begin{equation}
\begin{split}
\mu_{i, \sigma} &= f_{a}(E_i),  \\
\mu_{i, \pi} &= f_{a}(E_i)\sin^2\theta + f_{c}(E_i) \cos^2\theta, \\
\mu_{f, \sigma} &= f_{a}(E_f), \\
\mu_{f, \pi} &= f_{a}(E_f)\sin^2(\Omega - \theta) + f_{c}(E_f)\cos^2(\Omega - \theta).
\end{split}
\end{equation}
For LCO100 the absorption coefficients can be approximated as,
\begin{equation}
\begin{split}
\mu_{i, \sigma} &= f_{a}(E_i),  \\
\mu_{i, \pi} &= f_{c}(E_i)\sin^2\theta + f_{a}(E_i) \cos^2\theta, \\
\mu_{f, \sigma} &= f_{a}(E_f), \\
\mu_{f, \pi} &= f_{c}(E_f)\sin^2(\Omega - \theta) + f_{a}(E_f)\cos^2(\Omega - \theta).
\end{split}
\end{equation}

\begin{figure*}
\centering
\includegraphics[width=\linewidth]{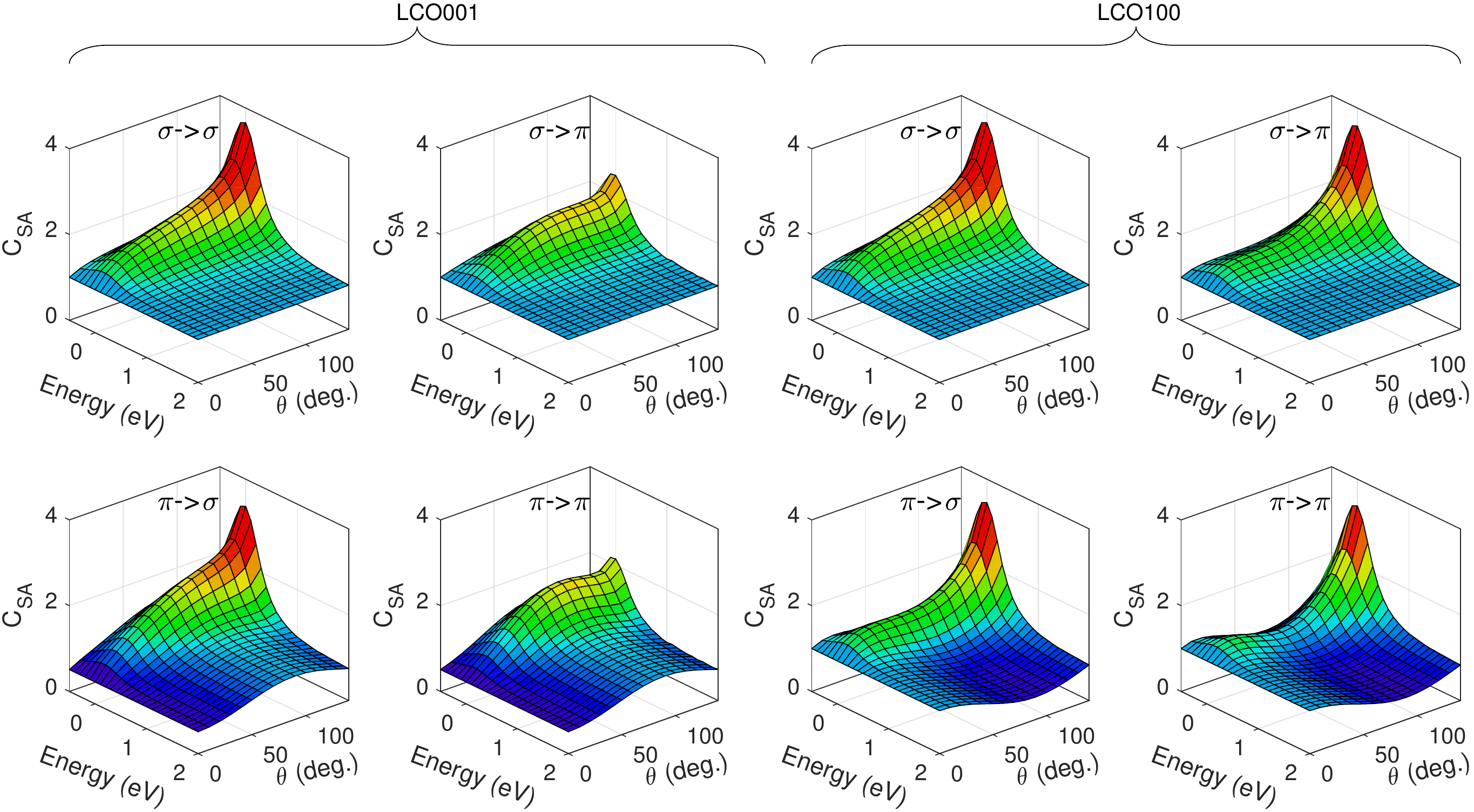}
\caption{The calculated self-absorption factor, $C_{\mathrm{SA}}$ for excitations measured in LCO as a function of excitation energy relative to Cu $L_3$ edge, incident angle $\theta$ and polarisation.}
\label{fig:norm_self_absorption_factors}
\end{figure*}

Taking these factors into account, Fig. \ref{fig:norm_self_absorption_factors} shows the energy, angle and polarisation dependence of the self-absorption factor, $C_{\mathrm{SA}}$. A peak in self-absorption is seen close to the elastic position which is most pronounced at large $
\theta$. However, it is clear that there is significant variation in the extent of self-absorption depending on the experimental setup.

\begin{figure*}
\centering
\includegraphics[width=\linewidth]{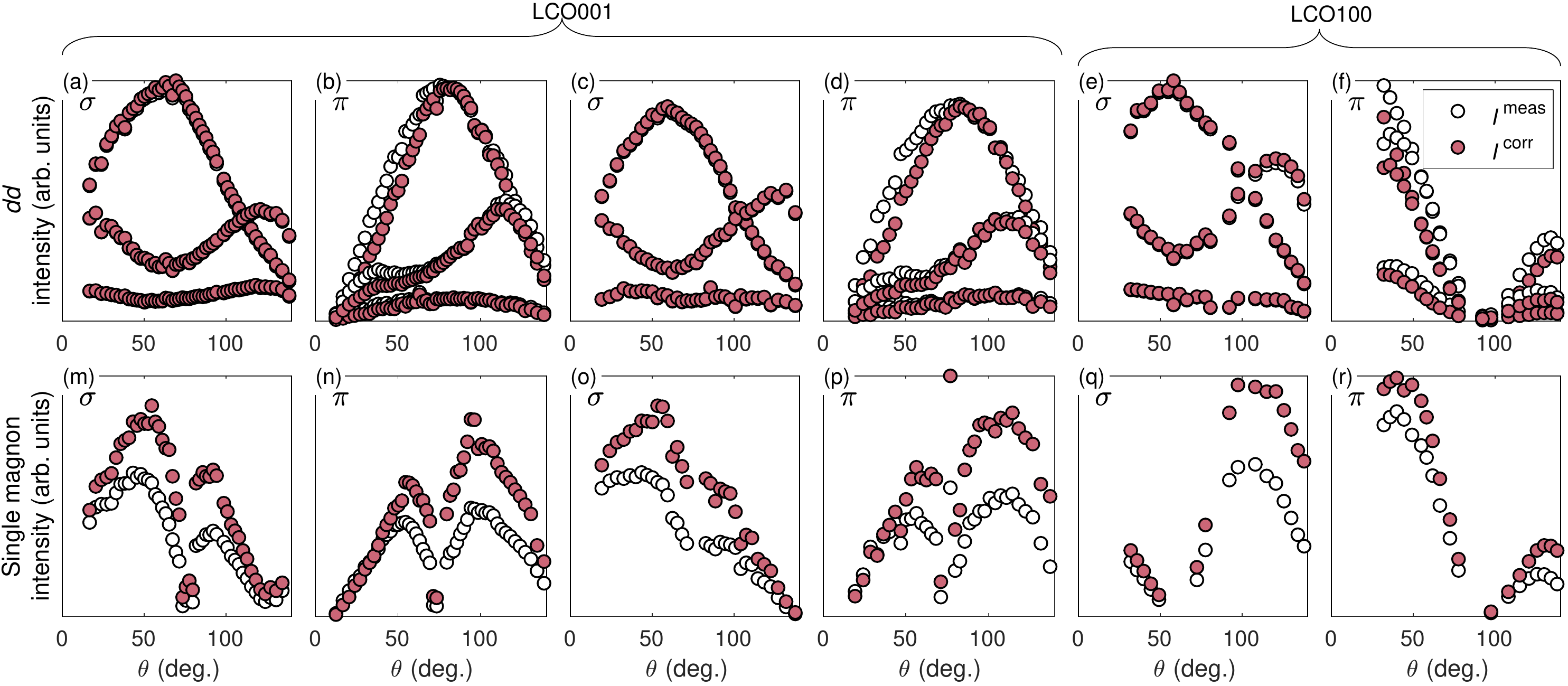}
\caption{RIXS intensity before and after applying the self-absorption correction. Measured data are shown in white and corrected data in red. The effect is shown for $dd$ excitations in panels (a-f) and in single magnons in panels (g-l).
\label{fig:excitation_intensity_corrected}}
\end{figure*}

Fig. \ref{fig:excitation_intensity_corrected} shows the result of applying the self-absorption correction to the different excitations that were measured. Panels (a-f) show the $dd$ excitations intensity before and after the self-absorption correction. As shown in Fig. \ref{fig:norm_self_absorption_factors}, at the energy where 
the $dd$ excitations occur, $\simeq 2$\;eV, the self-absorption factor is relatively low, therefore the corrected intensity is not significantly changed. For $dd$ excitations, we assume that the polarisation of the scattered photons is unchanged ($\sigma\rightarrow\sigma$ or $\pi\rightarrow\pi$). This is an approximation that does not account for the complexity of the $dd$ excitations reported in reference \cite{fumagalli2019} but as the total self-absorption at this energy is so small, this approximation works well enough for our purposes.

At low energy, the self-absorption is much greater and polarisation-dependence of the scattered light also becomes more significant. Fig. \ref{fig:excitation_intensity_corrected} (g-l) show the intensity of the single magnon excitations before and after the self-absorption correction. Here we assume the photon polarisation is flipped as a result of the excitation ($\sigma\rightarrow\pi$ or $\pi\rightarrow\sigma$). This assumption is justified by RIXS measurements performed with polarisation analysis such as that by Peng \textit{et al.} \cite{peng2018} and Fumagalli \textit{et al.} \cite{fumagalli2019}. Following this assumption, the single magnon intensity is seen to be significantly altered by the self-absorption effects.

\section{crystal field theory multiplet calculations}
\label{apx:CFT}
We computed the single-ion $dd$ orbital and spin-flip excitations using crystal field theory implemented in the Quanty package \cite{haverkort2012multiplet,haverkort2016quanty}. We define the ground and excited states and introduce interactions tensors between them. In both states the Cu$^{2+}$ basis has D$_{4h}$ symmetry with five orbitals, $d_{x^2-y^2}$, $d_{z^2}$, $d_{xy}$ and $d_{xz/xy}$. The energies of these orbitals are obtained through fittings showing good consistency with previous RIXS measurements \cite{moretti2011}. We summarise the values in Table \ref{table:I}.

\begin{table}
\begin{center}

\begin{tabular}{ c c c c c c } \\
\hline\hline
& $d_{x^2-y^2}$ (eV) & $d_{z^2} $(eV) & $d_{xy}$ (eV)& $d_{xz}$(eV)& $d_{yz}$(eV)\\
 \hline
 &0&1.5&1.8&2.1&2.1\\
 \hline\hline
\end{tabular}
\end{center}

\caption{Values of the orbital energies in LCO relative to the $d_{x^2-y^2}$ orbital. Parameters are obtained from RIXS measurements in reference \cite{moretti2011} and used in the single-site multiplet crystal field theory calculations of the local interactions. \label{table:I} }
\end{table}

Three interaction tensors act on these bases. The first, $U_{dd}$ describes the Coulomb repulsion between the $3d$ electrons. In Quanty, $U_{dd}$ is expanded as a sum of spherical harmonics and is defined with three Slater integrals, $F_{dd}^0$, $F_{dd}^2$ and $F_{dd}^4$. Additional Coulomb interaction between the $2p$ core hole and $3d$ electrons are defined by $F_{pd}^0$ and $F_{pd}^2$. The second interaction tensor defines the exchange-Coulomb interaction between the $2p$ core hole and $3d$ electrons in the intermediate state and are given by Slater integrals $G^1_{pd}$ and $G^3_{pd}$. The third interactions is spin-orbit coupling which is defined for the ground state and the core-hole state, $\xi_{d}$ and $\xi_{2p}$, respectively. To simulate the local spin-flip excitations, an inter-atomic exchange integral $H_{exch}$ is introduced. The parameters used here are given in Table \ref{table:II}.

\begin{table}
\begin{center}
\begin{tabular}{ c c c c c c c c c c} \\
 \hline\hline
&  $F_{dd}^0$&$F_{dd}^2$ & $F_{dd}^4$&$F_{pd}^0$  &$F_{pd}^2$ & $G_{pd}^1$ &$G_{pd}^3$ &  $\zeta_d$ & $\zeta_{2p}$ \\
 \hline
 $2p^6 3d^9$ & 0 &  12.854 & 7.980 & & & & & 0.102 &  \\
 $2p^5 3d^{10}$ & 0 & 13.611 & 8.457 & 0 & 8.177 & 6.169 & 3.510 & 0.124 & 13.498 \\ 
 \hline\hline
\end{tabular}
\end{center}

\caption{Parameters used in the single-site multiplet crystal field theory calculations of the local interactions in LCO. Parameters are from reference \cite{haverkort2005}. \label{table:II}}
\end{table}

The calculated RIXS intensity, $R(\boldsymbol{\epsilon},\boldsymbol{\epsilon}^{\prime}, \mathbf{k}, \mathbf{k}^{\prime})$, show strong energy, polarisation and angular dependences. For the high-energy $dd$ orbital excitations, spectra were fitted with 6 pseudovoigt functions which were integrated to yield $R_{\mathrm{orbital}}$ for each orbital. For the low-energy spin-flip excitations, the component was fitted with a damped-harmonic oscillator function which was integrated to yield $R_{\mathrm{spin}}$.

\end{document}